\documentclass[journal]{IEEEtran}
\usepackage{amsfonts}
\usepackage{graphicx}
\usepackage{algorithm}
\usepackage{algorithmic}
\usepackage{subfig}
\usepackage{amssymb}
\usepackage{amsmath}
\usepackage{setspace}
\usepackage{xcolor}
\IEEEoverridecommandlockouts
\begin{document}
\title{Dynamic Channel Access and Power Control in Wireless Interference Networks via Multi-Agent Deep Reinforcement Learning
\thanks{The material in this paper was presented in part at the IEEE Vehicular Technology Conference (VTC)-Fall, Honolulu, HI, Sep. 2019. }}

\author{Ziyang Lu, Chen Zhong and M. Cenk Gursoy \thanks{The authors are with the Department of Electrical
			Engineering and Computer Science, Syracuse University, Syracuse, NY, 13244. (e-mail: zlu112@syr.edu, czhong03@syr.edu, mcgursoy@syr.edu).}
	}

\maketitle

\thispagestyle{empty}

\begin{abstract}
Due to the scarcity in the wireless spectrum and limited energy resources especially in mobile applications, efficient resource allocation strategies are critical  in wireless networks. Motivated by the recent advances in deep reinforcement learning (DRL), we address multi-agent DRL-based joint dynamic channel access and power control in a wireless interference network. We first propose a multi-agent DRL algorithm with centralized training (DRL-CT) to tackle the joint resource allocation problem. In this case, the training is performed at the central unit (CU) and after training, the users make autonomous decisions on their transmission strategies with only local information. We demonstrate that with limited information exchange and faster convergence, DRL-CT algorithm can achieve 90\% of the performance achieved by the combination of weighted minimum mean square error (WMMSE) algorithm for power control and exhaustive search for dynamic channel access. In the second part of this paper, we consider distributed multi-agent DRL scenario in which each user conducts its own training and makes its decisions individually, acting as a DRL agent. Finally, as a compromise between centralized and fully distributed scenarios, we consider federated DRL (FDRL) to approach the performance of DRL-CT with the use of a central unit in training while limiting the information exchange and preserving privacy of the users in the wireless system. Via simulation results, we show that proposed learning frameworks lead to efficient adaptive channel access and power control policies in dynamic environments.
\end{abstract}
\begin{IEEEkeywords}
dynamic channel access, power control, multi-agent reinforcement learning, distributed learning, federated learning, wireless interference networks.
\end{IEEEkeywords}

\IEEEpeerreviewmaketitle

%  I. Introduction%%%%%%%%%%
\section{Introduction}
Radio spectrum has become a precious resource in wireless communications due to the inherent scarcity of the natural spectrum and the increase in the number of wireless devices \cite{chen2013stochastic}. Meanwhile, power control has also become critical in reducing the interference among wireless devices. These have led to intensive studies on the efficient spectrum access and power control/allocation policies in wireless networks. Most prior studies on wireless resource allocation are based on optimization frameworks and the corresponding algorithmic strategies. For instance, various approaches for dynamic spectrum access are surveyed in \cite{4205091} and \cite{4796930}. An iterative approach for wireless power allocation called as weighted minimum mean square error (WMMSE) is proposed in \cite{shi2011iteratively}. Another iterative algorithm based on fractional programming (FP) is employed in \cite{shen2018fractional} for power allocation. These algorithms can lead to efficient resource allocation strategies. However, they are mainly model-based approaches and require complete observation of the entire wireless system, which becomes intractable with the growing diversity and density of the networks \cite{zhang2019deep}.

Motivated by the success of machine learning (ML) including deep learning (DL) and reinforcement learning (RL), there has recently been increasing interest in deploying learning-based approaches for solving wireless resource allocation problems. A survey of recent results on the application of deep learning algorithms in different layers (e.g., physical layer, data link layer, and network layer) of intelligent wireless networks is provided in \cite{mao2018deep}.

In this paper, we jointly address dynamic channel access and power control in a wireless network via multi-agent deep reinforcement learning (DRL). In particular, we consider a wireless interference network with multiple transmitter-receiver pairs. Multiple frequency bands are available for transmission, and each transmitter is equipped with a DRL agent to make decisions on which channel to access for transmission and what power level to use. The main contributions of this paper can be listed as follows:

\begin{itemize}
	\item A DRL framework with centralized training (DRL-CT) is proposed to tackle the joint resource allocation problem in a wireless network. By setting different reward functions, the algorithm can be used to address specific objectives such as sum-rate maximization and proportional fairness. After the training, each user (as a DRL agent) in the network can make autonomous decisions on its transmission policy using only local information.
	\item The proposed DRL-CT algorithm is evaluated considering the following aspects: Performances with different reward functions are compared and analyzed, improvements achieved by introducing power control in addition to dynamic channel access are demonstrated, and the performance in a dynamic environment is identified.
	\item Distributed multi-agent reinforcement learning in which each user is capable of performing training and distributed decision making is proposed and studied. Motivated by the challenges in fully distributed deep reinforcement learning (DDRL) to keep high performance levels during testing in dynamic environments, a federated deep reinforcement learning (FDRL) algorithm is also proposed to imitate DRL-CT algorithm while reducing the communication cost and ensuring privacy among the users in the training phase.
\end{itemize}

The remainder of this paper is organized as follows: First, related prior studies are described in Section II. The system model is presented in Section III. We introduce the proposed DRL-CT framework, conduct the analysis and provide simulation results in Section IV. Subsequently in Section V, DDRL and FDRL algorithms are proposed for performing the learning in a distributed manner and the corresponding simulations results are provided. Finally, conclusions are drawn in Section VI.

% II. Related Work%%%%%%%%%%
\section{Related Work}

\subsection{Resource Allocation with Machine Learning}
Machine learning algorithms have already been deployed to address several resource allocation problems. In \cite{sun2017learning}, a deep learning framework is proposed for solving the power allocation problem and it is shown that a similar performance as that of the WMMSE algorithm can be achieved. The work in \cite{ye2018deep} addressed spectrum access and power allocation problem in vehicle-to-vehicle (V2V) communication via deep reinforcement learning. In \cite{nasir2019multi}, deep reinforcement learning is applied to determine distributed power allocation policies in wireless networks. It is demonstrated that comparable performance to that achieved by the WMMSE algorithm can be attained while reducing the execution time by approximately one half. Another learning-based approach \cite{wang2018deep} has addressed the dynamic spectrum access problem in wireless networks. All of the above works have shown the capability of deep reinforcement learning in solving resource allocation problems. For other related studies, we refer to surveys in  \cite{zhang2019deep} and \cite{luong2019applications}.

Aside from the aforementioned prior work, studies relatively closer to our work in this paper are conducted in \cite{naparstek2017deep} and \cite{naderializadeh2020resource}. In particular, in \cite{naparstek2017deep}, a deep reinforcement learning framework is proposed for dynamic spectrum access. In the work, the proposed DRL framework is able to maximize different objectives in a wireless network by setting different reward functions during the training. It is worth noting that the framework proposed in \cite{naparstek2017deep} dramatically reduces the amount of information exchange by proposing a novel state structure. In our work, different from \cite{naparstek2017deep} and \cite{wang2018deep}, we also consider power allocation in addition to spectrum access. In particular, we seek to find an optimal joint policy for both power allocation and spectrum access via DRL. Besides, we further conduct the proposed DRL framework in a distributed manner. In another recent work \cite{naderializadeh2020resource}, the authors proposed a multi-agent deep reinforcement learning framework to jointly tackle the user scheduling and power allocation problem, where the agents share their local information with their neighbours with a certain delay. Different from the work in \cite{naderializadeh2020resource}, we consider that the agents (i.e. transmitters) can be isolated and make decisions autonomously after the convergence of the learning algorithm.

\subsection{Federated Learning and Federated Reinforcement Learning}
With the emergence of powerful edge nodes and advances in edge computing, it has become desirable to conduct the learning algorithms in a distributed manner.  Federated Learning (FL) is a decentralized learning algorithm which is first proposed in \cite{mcmahan2017communication}. Using relatively few communication rounds among the decentralized agents, FL based on federated averaging (FedAvg) algorithm is able to train high-quality global models. FL enables each agent to train its model locally without the need of uploading its local data to the center node, and hence ensures privacy and reduces latency.

Recently, FL has been employed in solving wireless resource allocation problems. In \cite{samarakoon2018federated}, FL algorithm is considered for joint resource allocation in V2V networks. In our work, different from \cite{samarakoon2018federated}, we apply the FedAvg algorithm to the deep reinforcement learning algorithm and propose a federated DRL (FDRL) framework. In the literature, there exist recent studies on federated reinforcement learning. For instance, a novel federated reinforcement learning algorithm that is not based on FedAvg algorithm is proposed in \cite{DBLP:journals/corr/abs-1901-08277}. Federated reinforcement learning for fast personalization in video games is studied in \cite{nadiger2019federated}. To the best of our knowledge, our work in this paper is one of the first studies, applying this FDRL framework to address wireless resource allocation problems.

% III. System Model%%%%%%%%%%%%%%%
\section{System Model}
\subsection{Interference Channel Model}
\begin{figure}[!t]
\centering
\includegraphics[width=2.5in]{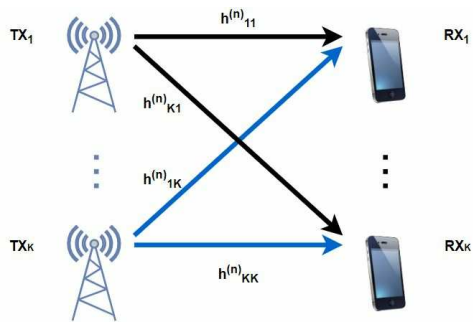}

\caption{Wireless interference network model}
\label{model}
\end{figure}

In the considered network model, we assume that there exist $K$ single-antenna transceiver pairs (i.e., $K$ transmitters and $K$ corresponding receivers) and $N_c$ available wireless channels (or equivalently frequency bands). Each transmitter-receiver pair can use a single channel at a time but multiple users can occupy the same channel, leading to interference. Let $h_{kk}^{\left ( n \right )}\in \mathbb{C}$ denote the direct-link fading coefficient of the transceiver pair $k$ in channel $n$ (for $k = 1,\ldots, K$ and $n = 1, \ldots, N_c$) and let $h_{kj}^{\left ( n \right )}\in \mathbb{C}$ denote the interference-link fading coefficient from transmitter $j$ to receiver $k$ in channel $n$.
Figure \ref{model} depicts the wireless interference network.

It is assumed that $h_{ij}^{\left (n  \right )}\sim \mathcal{CN}\left ( 0,1 \right )$, i.e., the fading coefficients of the channels are circularly symmetric complex Gaussian (CSCG) distributed with zero mean and unit variance.
%We further assume that the fading coefficients in different channels are independent of each other.

Throughout the paper, we generally refer to the transmitters in the considered wireless interference network as users.
The maximum transmission power of each user is limited by $P_{\max}$. Therefore, the transmit power of user $k$ satisfies $ 0\leq P_{k}\leq P_{\max} $.

Each user can be seen as a DRL agent and it uses a neural network to make decisions. For transmission, user $k$ needs to select a policy $\pi ^{k}$ for both power level and channel selection, i.e. $\pi ^{k} = \left ( \pi _{c}^{k},\pi _{p}^{k} \right )$, where $\pi _{c}^{k}\in \left \{ 1,2,3\cdots N_c \right \}$ is the channel selection policy (each user will select one out of $N_c$ channels) and $\pi _{p}^{k}\in \left [ 0,P_{\max} \right ]$ is the power selection policy for transmission. Our objective is to find the optimal policy $\pi = \left ( \pi ^{1},\pi ^{2}\cdots \pi ^{K} \right )$ for different utility objectives in the wireless network. The utility of user $k$ (when transmitting over channel $n$) is generally a function of its received signal-to-interference-plus-noise ratio (SINR), which can be formulated as
\begin{equation}
\text{SINR}_{k}= \frac{P_{k}\left | h_{kk}^{\left ( n \right )} \right |^{2}}{\sum_{j\neq k}^{ }P_{j}\left | h_{kj}^{\left ( n \right )} \right |^{2}+\sigma _{k}^{2}},
\end{equation}
where $\sigma _{k}^{2}$ is the variance of the Gaussian noise at receiver $k$, and the superscript $(n)$ denotes that the transmission of user $k$ occurs over channel $n$ chosen by the channel selection policy $\pi_c^k$. In the above formulation,  transmission powers $P$ are determined by the power selection policies of users.

\subsection{Dynamic Channel Model}\label{subsec:dynamicchannel}
After the training, DRL algorithms are tested in both static and dynamic environments to evaluate their robustness. In particular, in dynamic environments, in order to model the time-varying channel fading coefficients, Jakes' model is used to express the small-scale Rayleigh fading as a first-order Gaussian-Markov process as follows:
\begin{equation}
h_{ij}^{(n)(t+1)} = \rho h_{ij}^{(n)(t)} + e_{ij}^{(n)(t)}
\label{Vary}
\end{equation}
where $h_{ij}^{\left ( n \right )\left ( t \right )}$ denotes the $n^{th}$ channel's gain at time $t$. $e_{ij}^{\left ( n \right )\left ( t \right )}$ is an independent CSCG with mean zero and variance $(1-\rho^2)$, i.e., $\mathcal{CN}\left ( 0,1-\rho ^{2} \right )$. The correlation is $\rho = J_{0}\left ( 2\pi f_{d}T_{v} \right )$ where $J_{0}$ is the zeroth-order Bessel function of the first kind, $f_{d}$ is the maximum Doppler frequency and $T_{v}$ is the time interval over which the correlated channel variation occurs.

%We train the DRL algorithm over static channels with gains $h_{ij}^{(n)(0)}$. Then in the test phase, the channels start to vary according to (\ref{Vary}). Numerical results show that the DRL algorithm is capable of adapting to dynamic channels during the test, even if it is trained over static channels.

\subsection{Information Exchange} \label{subsec:informationexchange}
We assume that a certain exchange of information occurs between the transmitter and its corresponding receiver. In this paper, we consider the following feedback signal:
\begin{equation}
f_{k}\left ( t \right )=\left\{\begin{matrix}
n & \: \:  nT \le \text{SINR}_{k} < \left ( n+1 \right )T \\ 0
 & \: \: \text{SINR}_{k} < T
\end{matrix}\right.
\end{equation}
where $n$ denotes the integer feedback (requiring only finite number of bits in its binary representation), depending on the SINR levels, and  $T$ denotes the minimum required level for acceptable performance that depends on the quality of service requirements.

Assume that $f_k(t)$ is an integer, requiring at most $M$ bits in its binary representation. This implies that receiver $k$ quantizes the received $\text{SINR}_{k}$ to $2^{M}$ levels and feed back the results (an $M$-bit indicator) to transmitter $k$.
Selecting the number of quantized levels is a trade-off between performance and bandwidth/energy consumption during the transmission of the feedback signal.

We note that information exchange between different transmitters is not required in the proposed approach because each user can make its own decision based on only local information, i.e., their own transmission history and the feedback signal $f_{k}\left ( t \right )$ mentioned above. The state of each user is formed by its transmission history and $f_k(t)$ and details of state structure is illustrated in Section IV. The requirement of information exchange at the receiver side depends on the reward function selected for reinforcement learning. Details will be discussed in Section IV.

\subsection{Benchmark WMMSE Algorithm}
In \cite{sun2017learning}, weighted minimum mean square error (WMMSE) algorithm is used for power allocation to maximize the sum-rate in a wireless network. In this paper, WMMSE is modified for resource allocation in order to establish a benchmark on the performance in terms of the sum-rate. When only power allocation is considered, the sum-rate maximization problem can be formulated as
\begin{equation}
\begin{gathered}
\max_{p_{1},...,p_{k}}{\sum_{k=1}^{K}\log\left ( 1+\frac{p_{k}\left | h_{kk} \right |^{2}}{\sum_{j\neq k}^{ }p_{j}\left | h_{kj} \right |^{2}+\sigma _{k}^{2}} \right )} \\\\
s.t.\;\; 0\leq p_{k}\leq P_{\max}, \forall k = 1,2,...,K. \label{problem}
\end{gathered}
\end{equation}
Problem (\ref{problem}) is in general a challenging problem. In the WMMSE algorithm, the problem is described over a higher dimensional  space by introducing new variables $\text{mmse}_{k} = 1/\left ( 1+\text{SINR}_{k} \right )$, making the problem solvable. $\text{mmse}_{k}$ is known as the MMSE-SINR equality \cite{verdu1998multiuser}.

As noted in \cite{sun2017learning},  the sum-rate maximization problem (\ref{problem}) is then equivalent to the following weighted MSE minimization problem:
\begin{equation}
\begin{gathered}
\min _{\left \{ w_{k}, u_{k},v_{k} \right \}_{k=1}^{K}} \sum_{k=1}^{K}\left ( w_{k}e_{k} - \log\left ( w_{k} \right ) \right )\\\\
s.t. \; \;0\leq v_{k}\leq \sqrt{P_{\max}}, k = 1,2,...,K\\
e_{k} = \left ( u_{k}\left | h_{kk} \right |v_{k} - 1 \right )^{2} + \sum_{j\neq k}\left ( u_{k}\left | h_{kj} \right |v_{k}\right )^{2} + \sigma_{k}^{2}u_{k}^{2}. \label{newproblem}
\end{gathered}
\end{equation}

Then the above problem can be solved iteratively. In each iteration, only one set of variables is optimized while the other variables are fixed. Details can be found in \cite{shi2011iteratively} and \cite{sun2017learning}.

Because both spectrum access and power allocation are considered in our work, we use the WMMSE method in the following manner: First, each user selects a channel for transmission and the channel selection policies for all $K$ users are represented by $\pi _{c} = \left ( \pi _{c}^{1},  \pi _{c}^{2}, \cdots \pi _{c}^{K} \right )$. Then, WMMSE algorithm is used in each channel to find the near-optimal power allocation, i.e. $\pi _{p}^{\ast} = \left ( \pi _{p}^{1},  \pi _{p}^{2}, \cdots \pi _{p}^{K} \right )$. It is straightforward that the policy $\pi = \left(\pi _{c},\pi _{p}^{\ast}\right) $ has the maximal sum-rate for given channel selection policies in $\pi_{c}$. The procedure is repeated for all the possible $\pi_{c}$ (i.e., exhaustive search is performed). Then the near-optimal resource allocation policy $\left( \pi_{c}^{\ast}, \pi_{p}^{\ast}\right)$ for  maximizing the sum-rate in the network can be determined. Detailed steps are provided in Algorithm 1 below.

\begin{algorithm}[!ht]
\caption{WMMSE Algorithm for resource allocation}
\begin{spacing}{0.8}
\begin{algorithmic}[1]
\FOR{each scheme of channel selection}
\STATE{Initialize $v_{k}^{0}\in \left [ 0,\sqrt{P_{k}} \right ],\forall k$;}
\STATE{Compute $u_{k}^{0} = \frac{\left | h_{kk} \right |v_{k}^{0}}{\sum_{j = 1}^{K}\left | h_{kj}\right |^{2}\left ( v_{j}^{0} \right )^{2}+\sigma_{k}^{2}}, \forall k$;}
\STATE{Compute $w_{k}^{0} = \frac{1}{1-u_{k}^{0}\left | h_{kk} \right |v_{k}^{0}}, \forall k$;}
\FOR{iteration t = $0,1,2,...T_{max}$}
\STATE{Update $v_{k}$: $v_{k}^{t}=\left [ \frac{w_{k}^{t-1}u_{k}^{t-1}\left | h_{kk} \right |}{\sum_{j=1}^{K}w_{j}^{t-1}\left ( u_{j}^{t-1} \right )^{2}\left | h_{jk} \right |^{2}} \right ]_{0}^{P_{max}}, \forall k$;}
\STATE{Update $u_{k}$: $u_{k}^{t} = \frac{\left | h_{kk}\right |v_{k}^{t}}{\sum_{j=1}^{K}\left | h_{kj} \right |^{2}\left ( v_{j}^{t} \right )^{2}+\sigma _{k}^{2}},\forall k$;}
\STATE{Update $w_{k}$: $w_{k}^{t} = \frac{1}{1-u_{k}^{t}\left | h_{kk} \right |v_{k}^{t}}, \forall k$;}
\IF{Some stopping criteria is satisfied}
\STATE{break;}
\ENDIF
\ENDFOR
\STATE{Compute the sum-rate $s^{t} = \sum_{k = 1}^{K}\log\left(1+u_{k}^{t}\right)$};
\ENDFOR
\STATE{Compare the sum-rate of all the channel schemes obtained by the above WMMSE method. The largest one corresponds to the optimal scheme of resource allocation.}
\end{algorithmic}
\end{spacing}
\end{algorithm}

It can be seen from Algorithm 1 that WMMSE is centralized and requires full knowledge of the channel state information (CSI) including the direct and interference gains, which is not very feasible in practice .
On the other hand, by the DRL algorithm proposed in this paper, each user can make autonomous decisions for different objectives (sum-rate maximization or individual rate maximization) only by using its feedback signals  ($f_{k}\left ( t \right )$). The information exchange between the users is limited (especially after the DQN is trained), rendering the DRL algorithm scalable.

%IV. CDRL%%%%%%%%%%%%%%%%%%%%%%%%
\section{Deep Reinforcement Learning with Centralized Training for Joint Channel Access and Power Control}
In this section, we apply DRL with centralized training (DRL-CT) or more specifically employ deep Q-learning (DQN) and actor-critic to determine efficient policies to optimize certain objectives in wireless networks. We first provide the problem statement and then introduce the methodologies of deep Q-learning and actor-critic. Subsequently, we design our deep reinforcement learning-based agents for joint channel access and power control.

\subsection{Problem Statement}

In this work, the goal is to determine the joint channel access and power control policies that maximize a given network utility. Hence, we address a decision-making problem in a dynamic wireless environment, and utilize DRL to determine efficient policies. Specifically, in this setting, each user (transmitter) is a DRL agent that needs to select an available channel and a power level to transmit in order to maximize a specific utility function. We consider different utility functions (e.g., sum rate, sum log-rate, and individual rates) and provide more details in Section \ref{subsec:states-actions-rewards}. In particular, the reward formulations reflect the objective of the problem. In the considered setting, WMMSE or exhaustive search become infeasible especially for large-scale networks since WMMSE requires channel information and the complexity of exhaustive search increases exponentially. Therefore, the proposed DRL framework becomes an efficient strategy to address this dynamic control problem.

\subsection{Deep Q-Learning}
In Q-learning, the agents learn an optimal policy by interacting with the environment and obtaining rewards for the actions taken. Generally, the aim of reinforcement learning is to learn a policy that can achieve the maximum discounted sum reward, i.e. can maximize
\begin{equation}\label{eq:reward}
R=\sum_{i=0}^{\infty }\gamma ^{i}r_{i+1}
\end{equation}
where $\gamma$ is the discount factor and $r_{i}$ is the instantaneous reward at time $i$.

 Let $Q^{\pi }\left ( s, a \right )$ denote the action-value function (i.e., the Q-value function), which describes the expected discounted sum of rewards the agent can obtain by taking action $a$ in state $s$ following action policy $\pi$. The agent's goal is to learn to choose the optimal action value $Q^{\ast }\left ( s, a \right )= \max _{\pi }Q^{\pi}\left ( s, a \right )$ and determine the optimal policy. The estimated $Q$ values are stored in a lookup table called $Q$-table which maps the states to the actions. The $Q$ values are updated according to the equation below:
\begin{equation}
Q\left ( s, a \right )=Q\left ( s, a \right )+\alpha \left ( r+\gamma  \max _{a'}Q\left ( {s}',a' \right )-Q\left ( s, a \right ) \right )
\end{equation}
where $\alpha$ is the learning rate and $s'$ is the next state following $s$.

We assume that the users select the actions based on the $\varepsilon$-greedy method with a small decreasing value of $\varepsilon$. In this scheme, the agent chooses the action with the largest $Q(s, a)$ value in state $s$ with probability $1-\varepsilon$ and selects an action randomly with probability $\varepsilon$.

Classical Q-learning performs well with small state and action spaces. However, in resource allocation problems, the action space can be large because each user has to select its power level as well as the transmission channel. In such cases, deep Q-learning performs better and becomes preferable \cite{mnih2015human}. %In \cite{mnih2015human}, deep reinforcement learning has been used in training computers to play Atari games for which strong performance levels have been demonstrated.
In the algorithm of deep Q-learning, a neural network is employed in place of the Q-table for mapping the states to the actions. In certain applications, dueling DQN \cite{wang2016dueling} can be used to reduce the occurrence of states which lead to poor performance regardless of the taken actions, and with this, the performance can be improved.

In DQN, parameters $\{w\}$ of the neural network are updated by the method of experience replay and backpropagation. Each agent will store its experience tuple $(s, a, r, s^{'})$ in a memory replay buffer. A mini batch $\mathbb{M}$, which is a randomly sampled subset of the memory replay buffer, is used to calculate the gradient for updating $w$ according to the rule of backpropagation. We can express the gradient averaged over the mini batch as
\begin{equation}
g =\frac{1}{\left |\mathbb{M}  \right |}\sum_{i=1}^{\left |\mathbb{M}  \right |}\phi(sample_{i} \in \mathbb{M}).
\label{BackProp}
\end{equation}
where $\phi$ denotes the gradient computed using the $i^{th}$ sample from the mini batch and $\left |\mathbb{M}  \right |$ is the cardinality of $\mathbb{M}$.
After calculating the gradient, $w$ will be updated as
%\label{CDRLupdate}
\begin{equation}
w = w - \alpha g = w - \alpha \frac{1}{\left |\mathbb{M}  \right |}\sum_{i=1}^{\left |\mathbb{M}  \right |}\phi(sample_{i} \in \mathbb{M})
\label{MiniBatch}
\end{equation}
where $\alpha$ denotes the learning rate.

\subsection{Actor-Critic}

With the common objective, the actor-critic agent also aims at finding an optimal action policy $\pi$ to maximize the reward given in (\ref{eq:reward}). Different from the value-based deep Q-learning, the actor-critic framework is a policy gradient reinforcement learning algorithm that consists of two neural networks: the actor neural network is used for action selection, and the critic neural network is used to evaluate the action policy. Below we briefly describe how these two neural networks collaborate to achieve the maximum reward.

\emph{Actor:} The actor is employed to explore the action policy $\pi$, that maps the agent's observation $s$ to the action space $\mathcal{A}$:
\begin{equation}
\pi_{\theta}(s) : s \rightarrow \mathcal{A}.
\end{equation}
So the mapping policy $\pi_{\theta}(s)$ is a function of the observation $s$ and is parameterized by $\theta$. And the chosen action can be denoted as
\begin{equation}
a = \pi_{\theta}(s)
\end{equation}
where we have $a \in \mathcal{A}$. We use softmax function at the output layer of the actor network so that we can obtain the score of each action. The scores sum up to $1$ and can be regarded as the probabilities to obtain a good reward by choosing the corresponding actions.

\emph{Critic:} The critic is employed to estimate the value function $V(s)$. At time instant $t$, when the action $a(t)$ is chosen by the actor network, the agent will execute it in the environment and send the current observation $s(t) $ along with the feedback from the environment to the critic. The feedback includes the reward $r_t$ and the next time instant observation $s(t+1)$. Then, the critic calculates the TD (Temporal Difference) error:
\begin{equation} \label{eq:TDerror}
\delta_{t} = r\left( t \right) + \gamma V_{\mu}(s\left( t+1 \right)) - V_{\mu}(s\left( t \right))
\end{equation}
where $\gamma \in (0,1)$ is the discount factor.

\emph{Update:} The critic is updated by minimizing the least squares temporal difference (LSTD):
\begin{equation}
%\underset{V^*}{\text{minimize}} \quad (\delta^{\pi_\theta} )^2
V^* = \arg \min_{V_{\mu}} (\delta_{t} )^2
\end{equation}
where $V^*$ denotes the optimal value function.

The actor is updated by policy gradient. Here, we use the TD error to compute the policy gradient\footnote{In (\ref{eq:policygradient}), policy gradient is denoted by $\nabla_{\theta} J(\theta)$ where $J(\theta)$ stands for the policy objective function, which is generally formulated as the statistical average of the reward.}:
\begin{equation}\label{eq:policygradient}
\nabla_{\theta} J(\theta) = E_{\pi_{\theta} } [ \nabla_{\theta} \log \pi_\theta(s, a)  \delta_{t} ]
\end{equation}
where $\pi_\theta(s, a)$ denotes the score of action $a$ under the current policy. Then, the weighted difference of parameters in the actor at time $t$ can be denoted as $\Delta\theta_{t} = \alpha \nabla_{\theta_t} \log \pi_{\theta_t}(s\left( t \right), a\left( t \right)) \delta_{t}$, where $\alpha \in (0,1)$ is the learning rate. And the actor network $i$ can be updated using the gradient decent method:
\begin{equation}
\theta_{t+1} = \theta_t + \alpha \nabla_{\theta_t} \log \pi_{\theta_t}(s\left( t \right), a\left( t \right)) \delta_{t}.
\end{equation}

\subsection{Proposed Deep Reinforcement Learning Model} \label{subsec:states-actions-rewards}

As noted before, in this section, we assume that the DRL-based agents are initially trained centrally and each user will have the same copy of the neural networks. Once the DRL action policy converges, the central unit is taken away from the system and the users make decisions autonomously with the well-trained DQN or actor networks and local information. Details on the states, actions, and rewards are provided below.

\subsubsection{States}
\begin{figure}[!t]
\centering
\includegraphics[width=3.5in]{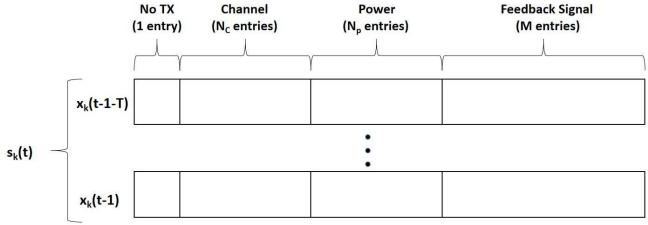}

\caption{State structure}
\label{state}
\end{figure}

Recall that we address joint channel access and power allocation. The observation $x_{k}\left ( t \right )$ of user $k$ at time $t$ is a vector of size $1+N_c+N_p+M$ \emph{with binary entries}, where $N_c$ denotes the number of available channels and $N_p$ denotes the number of power levels between  $P_{\min}$ and $P_{\max}$ (and we generally assume $P_{\min} = 0$). The set of available power levels can then be represented as $P=(P_1, P_2, ..., P_{N_p})$ in the ascending order, where $P_1 = P_{\min}$ and $P_{N_p}=P_{\max}$.  $M$ denotes the length of the binary representation of the feedback signal $f_{k}\left ( t \right )$. The first entry of the vector is an indicator of no transmission, i.e., there is no transmission if it is set to $1$. Otherwise, it is zero.  The following $N_c$ entries indicate the channel selected by user $k$. The next $N_p$ entries indicate the power level of the user. The last $M$ entries is the feedback signal $f_{k}\left ( t \right )$. For example, if user $k$ chooses channel $n$ and power level $P_i$ to transmit, the $\left (n+1 \right )^{th}$ entry and the $\left ( N_c+ i + 1 \right )^{th}$ entry will be set to 1. The last $M$ entries of the vector will be the binary representation of the feedback signal $f_{k}\left ( t \right )$. Other entries will remain 0. If the user chooses not to transmit, the first entry is set to 1 and all the other entries of the vector will be 0. The state $s_{k}(t)$ of the user $k$ at time $t$ is defined as the collection of observations in the previous $T$ time slots. The structure of the states is shown in Fig. \ref{state}.

The actions of the other users at time $t$ can only be partially observed via the feedback signal of user $k$. The learning will be efficient if the user can aggregate the past observations and figure out its internal state at time $t$. Hence, a long short term memory (LSTM) layer \cite{hausknecht2015deep} is used to realize this.
\subsubsection{Actions}
An action is the output of the action policies (DQN or the actor network in the actor-critic algorithm). The action is determined by considering a combination policy that consists of both the channel selection policy $\left ( \pi _{c}^{}\in{\left[1,N_c\right]} \right )$ and the power allocation policy $\left ( \pi _{p}^{}\in{\left[P_{\min},P_{\max}\right]} \right )$. Hence, the action  can be expressed as $a= \left ( \pi _{c} ,\pi _{p}\right )$. There are totally $N_c$ channels available and the transmit power of the users are quantized to $N_p$ levels, which results in $N_c\times N_p$ possible actions when transmission occurs\footnote{If $P_1 = P_{min} = 0$, the number of actions when transmitting is $N_c \times (N_p-1)$.} , and $1$ action in case of a no transmission. Therefore, there are in total $N_c\times N_p + 1$ possible actions for each user.
\subsubsection{Rewards}
How to set the reward depends on the objective within the network, i.e., the utility function to be maximized. In this paper, we consider three different rewards:
\begin{itemize}
	\item \emph{Individual transmission rate}: The reward of user $k$ at time $t$ denoted by $r_k(t)$ can be set as its transmission rate at time $t$, i.e.
\begin{equation}
r_{k}\left ( t \right ) = \log_{2}\left ( 1+\text{SINR}_{k}\left ( t \right ) \right ).
\end{equation}
The users will then attempt to competitively maximize their own transmission rates.
	\item \emph{Sum-rate}: The reward of all the users in time slot $t$ is set to be the sum-rate achieved at $t$, i.e.,
\begin{equation}
r\left ( t \right ) =\sum_{k=1}^{K}\log_{2}\left ( 1+\text{SINR}_{k} \left ( t \right )\right ).\label{reward_sum}
\end{equation}
In this scenario, all the users work cooperatively and decide their policy with the goal to maximize the sum-rate in the network.
	\item \emph{Proportional Fairness}: By setting the objective as maximizing the users sum log-rate is known as an approach to achieve proportional fairness \cite{kar2004achieving}. In this case, the reward at time $t$ is set as \begin{equation}
r\left ( t \right ) =\sum_{k=1}^{K}\log\left (\log_{2}\left (1+ \text{SINR}_{k}\left ( t \right ) \right ) \right ).
\end{equation}
 In this case, the users will work cooperatively to maximize this indicator of fairness.
\end{itemize}

We note that rewards are provided to the DRL agents only during the training period (unlike the quantized SINR feedback, described in Section \ref{subsec:informationexchange}, which is sent as local information from the corresponding receiver during both training and testing). If individual transmission rate is the reward, this transmission rate is either already known by each transmitter or it can be provided to each transmitting agent by the corresponding receiver in a distributed fashion. If sum rate or sum log-rate is the reward, some coordination may be needed. In particular, in centralized training, such information can be collected by the central unit and fed back to the users as the reward. In distributed training, sum rate information can be obtained by the DRL agents by listening all the feedback from the corresponding receivers. As another approach, the receivers or the transmitters can coordinate to learn the sum-rates. If these options are not available, DRL agents need to work with individual rate rewards.

In the simulation results in Section IV.E below, performance comparisons with different rewards will be provided.

\subsection{Workflow and Algorithm}

In the proposed algorithm in this section, the DRL model is centrally trained and then distributively used by the users in the wireless network:
\begin{itemize}
\item \emph{Training}: Training is performed in the central unit (CU). At the beginning of each time slot, transmitter $k$ in the network will choose a policy $\pi^{k} = \left( \pi_{c}^{k}, \pi_{p}^{k}\right)$ based on the $\epsilon$-greedy method and transmit simultaneously with the other users. After receiving the signal, the receivers will send a real-valued reward ($r_{k}\left(t\right)$ or $r(t)$) and the $M$-bit feedback $f_{k}^{t}$ back to their corresponding transmitters. Then, each transmitter gets access to its transmission experience, e.g., the state, action and reward described in Section \ref{subsec:states-actions-rewards}. Following this, all the users will upload their own experiences to the CU for updating the weights of the corresponding DRL model.
\item \emph{Usage}: When the action policy converges, each user will get a copy of the DRL model distributed from CU and make their own decisions autonomously with only the quantized SINR feedback. The inputs required for DRL agents is the history of actions and observations in the past several time slots, i.e. the local information. Since the transmission history is available locally, information exchange only happens when sending back the observation. Central training can be periodically repeated in case of significant variations in the channel gains.

\end{itemize}

The proposed deep reinforcement learning algorithm (with the specific operations of DQN and actor-critic agents described in Algorithms 3 and 4) is shown in Algorithm 2.

%\begin{algorithm}[!ht]
%\caption{CDRL Algorithm}
%\begin{spacing}{0.8}
%\begin{algorithmic}[1]
%\FOR{time-slot $t = 1,...,T_{max}$}
%\STATE{All the users are making decisions simultaneously.}
%\STATE{Each user $k$ will get access to its experience in the past T time-slot, i.e. $$s_{k}\left (t \right ) = \left ( \left \{ a_{k}\left ( i \right )_{i = t-T}^{t} \right \},\left \{ f_{k}\left ( i \right )_{i=t-T}^{t} \right \}\right ),$$ where $a_{k}(i)$ denotes the action taken by user $k$ in the time-slot $i$, $f_{k}(i)$ is the feedback signal. }
%\STATE{Each user selects an action $a$ with DQN and $\epsilon$-greedy method in the current state $s_{k}\left (t \right )$.}
%\STATE{After the transmissions, each user will be given a feedback signal and a reward $r$ for action $a$ in the state $s_{k}\left (t \right )$. }
%\STATE{$s_{k}\left (t \right ) = s_{k}\left (t + 1 \right )$.}
%\STATE{Each user uploads $\left ( s_{k}\left ( t \right ),a,r,s_{k}\left ( t + 1 \right ) \right )$ to the central unit.}
%\STATE{Update DQN with a mini-batch randomly selected from the memory replay buffer.}
%\ENDFOR
%\end{algorithmic}
%\end{spacing}
%\end{algorithm}

\begin{algorithm}[!ht]
	\caption{CDRL Algorithm}
	\begin{spacing}{0.8}
		\begin{algorithmic}[1]
			\STATE Call the \textbf{Initialization} function of the corresponding agent to establish the neural networks.
			\FOR{time-slot $t = 1,...,T_{max}$}
			\STATE{All the users are making decisions simultaneously.}
			\STATE{Each user $k$ will get access to its experience in the past $T$ time slots, i.e.
			$$s_k(t) = \{x_k(i)\}_{i=t-1-T}^{t-1}$$
			where $x_{k}(i)$ denotes the observation of user $k$ in the time-slot $i$.}
			\STATE{Each user calls the \textbf{Action Selection} function of the corresponding DRL-based agent to select an action $a$ with $\epsilon$-greedy method in the current state $s_{k}\left (t \right )$.}
			\STATE{After the transmissions, each user will be given a feedback signal and a reward $r$ for action $a$ in state $s_{k}\left (t \right )$. }
			\STATE{$s_{k}\left (t \right ) = s_{k}\left (t + 1 \right )$.}
			\STATE{Each user uploads $\{s_{k}\left (t \right ), a(t), r\left( t \right), s_{k}\left (t+1 \right ) \}$ tuple to the central unit and updates the corresponding agents by calling the \textbf{Update} functions of the corresponding agent.}
			\ENDFOR
		\end{algorithmic}
	\end{spacing}
\end{algorithm}

%$a\left( t \right) = \pi_{\theta}(s_{k}\left (t \right )| \theta)$
\begin{algorithm}
	\caption{DQN Agent}
	\label{alg:DQN}
	\begin{algorithmic}
		\STATE \textbf{Initialization:}
		\STATE Initialize the parameters of the DQN network $\pi_{\theta}(s\left (t \right ))$, parameterized by $\theta$.
			
		\STATE \textbf{Action Selection:}
		\STATE Receive the current observation $s_{k}\left (t \right )$ from the main function.
		\STATE With the observation, the DQN agent selects an action according to the decision policy (the current DQN network) $a\left( t \right) = \arg\max _{\pi }Q^{\pi}\left ( s, a \right )$.
		\STATE Return the selected action $a(t)$ to the main function.
		
		\STATE \textbf{Update:}
		\STATE Receive the $\{s_{k}\left (t \right ), a(t), r\left( t \right), s_{k}\left (t+1 \right ) \}$ tuple from the main function and store it into the memory replay buffer of the DQN agent.

		\STATE Update the DQN by minimizing the loss via the method of experience replay and backpropagation as shown in equations (\ref{BackProp}) and (\ref{MiniBatch}).

	\end{algorithmic}
\end{algorithm}

\begin{algorithm}
	\caption{Actor-Critic Agent}
	\label{alg:AC-SU}
	\begin{algorithmic}
		\STATE \textbf{Initialization:}
		\STATE Initialize the critic network $V_{\mu}(s\left (t \right ) )$ and the actor $\pi_{\theta}(s\left (t \right ))$, parameterized by $\mu$ and $\theta$ respectively.

		\STATE \textbf{Action Selection:}
		\STATE Receive the current observation $s_{k}\left (t \right )$ from the main function.
		\STATE With the observation, the actor-critic agent selects an action according to the decision policy (the current actor network) $a\left( t \right) = \pi(s_{k}\left (t \right )| \theta)$ w.r.t. the current policy
		\STATE Return the selected action $a_t$ to the main function.
	
		\STATE \textbf{Update\footnotemark:}
		\STATE Receive the $\{s_{k}\left (t \right ), r\left( t \right), s_{k}\left (t+1 \right ) \}$ tuple from the main function.
		\STATE Critic calculates the TD error: $ \delta_{t} = r\left( t \right) + \gamma V(s_{k}\left (t+1 \right )) - V(s_{k}\left (t \right )) $
		\STATE Update the critic by minimizing the loss: $\mathcal{L}(s_{k}\left (t \right ), a\left( t \right)) = (\delta_{t} )^2$
		\STATE Update the actor policy by maximizing the action value: $\Delta\theta_t = \alpha \nabla_{\theta_t} \log \pi_{\theta_t}(s_{k}\left (t \right ), a\left( t \right)) \delta_{t}$, $\alpha \in (0,1)$.

	\end{algorithmic}
\end{algorithm}
\footnotetext{When the function is called by the CDRL algorithm, the critic network is updated for every user, and the actor network is updated for only one randomly selected user in every iteration. When the function is called by the FDRL algorithm, both of the critic and actor networks are updated for every user in the group $\mathbb{G}$.}

\subsection{Simulation Results}
In this section, we initially describe the simulation setup and then provide detailed discussions on the simulation results.
%\textcolor{red}{Add simulation results of actor critic.}
\subsubsection{Simulation Setup}
It is assumed that there are $K = 6$ transceiver pairs (i.e., user pairs) and $N_c = 2$ channels in the wireless network. Without loss of generality, we consider this relative small-scale network in the experiments because the computation cost of the benchmark (exhaustive search and WMMSE) increases exponentially with the increase in the number of users $K$.

In the deep Q-network (DQN), we have one input layer, one LSTM layer with 20 neurons, a dueling network consisting of an advantage layer and a value layer with 10 neurons in each, and one output layer. The probability $\epsilon$ is assumed to uniformly decrease from 0.2 to 0.01 during the first 80 percent of the training time. Adam Optimizer is used to conduct the stochastic gradient descent to update the weights in the DQN. The DQN is trained for 400000 time-slots to ensure convergence. The other parameter values are listed in Table \ref{table1} below.

In the actor-critic DRL agent, the actor network is set to have the same structure as the DQN network with the learning rate of 0.0005. The critic network has one input layer, one fully-connected layer with 5 neurons and one output layer, with the learning rate of 0.001.

\begin{table}[!ht]
\renewcommand{\arraystretch}{1.3}

\caption{Experimental Parameters}
\label{table1}
\centering
\small
\begin{tabular}{|c||c|}
\hline Noise Power $\left ( \sigma ^{2} \right )$&1\\

\hline Number of Users $\left (K \right )$&6\\

\hline Number of Channels $\left (N_c \right )$&2\\

\hline Power Limit $\left (P_{max}\right )$&38dBm\\

\hline Power Levels $\left (N_p \right )$&5\\

\hline Length of the Feedback Signal $(M)$ & 10 bits\\

\hline Training Time $\left(T_{max}\right)$ &400000\\

\hline Exploration Time $\left(T_{e}\right)$ &320000\\

\hline Learning rate $\left (\alpha \right )$&0.001\\

\hline Discount Factor $\left (\gamma \right )$ & $0.9$\\

\hline Initial Exploration Probability $\left (\epsilon _{max}\right )$ &0.2\\

\hline Final Exploration Probability $\left (\epsilon _{min}\right )$ &0.01\\
\hline Optimizer &Adam\\

\hline
\end{tabular}
\end{table}

\subsubsection{Performance Results}
We generate $N_c$ matrices with the size of $K\times K$ independently as the channel fading coefficients according to a complex Gaussian distribution, i.e., $h_{ij}^{\left (n  \right )}\sim \mathcal{CN}\left ( 0,1 \right )$, at the beginning of the experiment.

\begin{figure}[!t]
\centering
\includegraphics[width=3.5in]{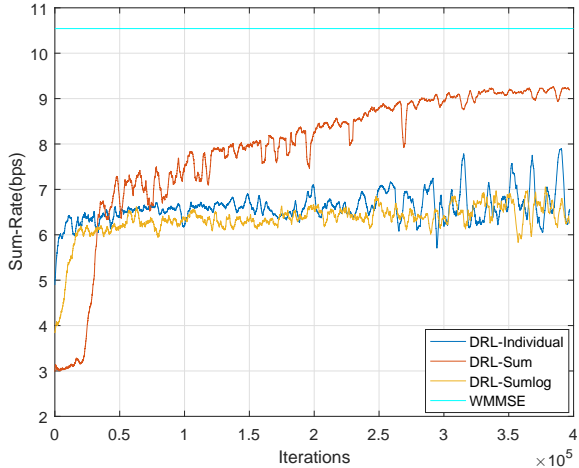}

\caption{Moving average sum-rate achieved with three different reward mechanisms in the proposed DQN algorithm.}
\label{sum}
\end{figure}

\begin{figure}[!t]
\centering
\includegraphics[width=3.5in]{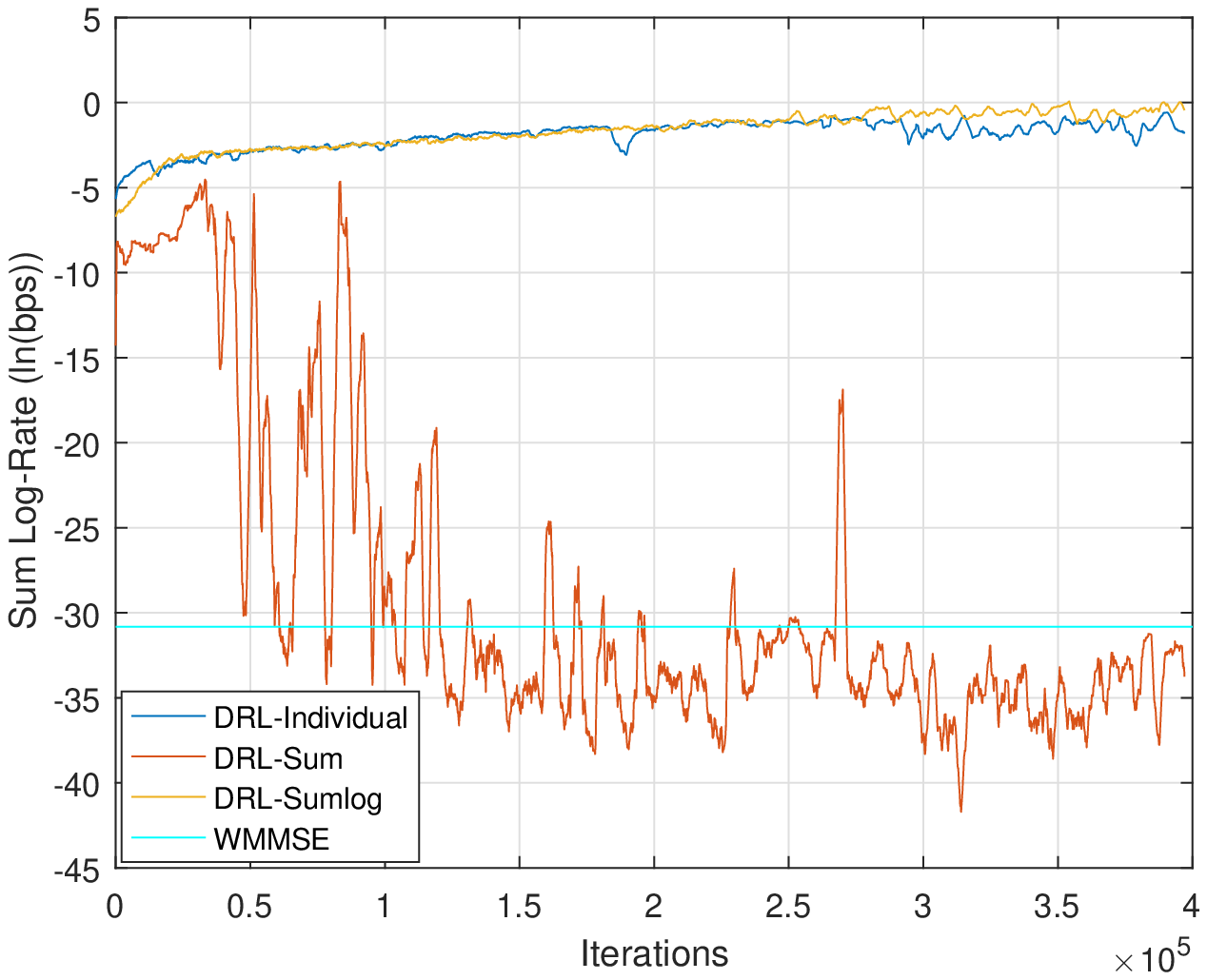}

\caption{Moving average sum log-rate achieved with three different reward mechanisms in the proposed DQN algorithm.}
\label{sumlog}
\end{figure}

Fig. \ref{sum} plots the moving average sum-rate in the wireless network achieved by the proposed DQN algorithm considering three different reward functions and Fig. \ref{sumlog} plots the corresponding moving average sum log-rate. In Fig.  \ref{sum}, it can be seen that the best performance in terms of the sum-rate is achieved when the sum-rate is selected as the reward function. And the sum-rate in this case is approximately 90\% of the sum-rate achieved by WMMSE and exhaustive search for optimal channel selection. Note that WMMSE requires full CSI while the actions of all the users and the CSI are not available in the proposed DQN algorithm. We also observe in the figure that using individual rate as the reward leads to some performance degradation. Finally, when sum-log rate is used as the reward, proportional fairness is addressed in the network and as a result the sum-rate performance degrades further.

In Fig. \ref{sumlog}, highest performance in terms of the sum log-rate is attained when sum log-rate is considered as the reward function. Using individual rate as the reward results in comparable sum log-rate levels due to the fact that competitive user behavior in this case leads to a certain level of proportional fairness. Note that this comparable performance in terms of the sum log-rate is achieved (in the case of individual rate reward) with local information feedback only. Hence, no information exchange between the users is required in the network, which makes the algorithm scalable and hence more practical in large-scale networks.

On the other hand, using sum-rate as the reward performs poorly in terms of fairness, and achieves the lowest sum log-rate levels. Note that when sum-rate is the reward, some users may be encouraged to not transmit or use less power for transmission, and this potentially leads to small or even negative contributions due to negative values of the logarithm of the small rates.

\begin{figure}[!t]
	\centering
	\includegraphics[width=3.5in]{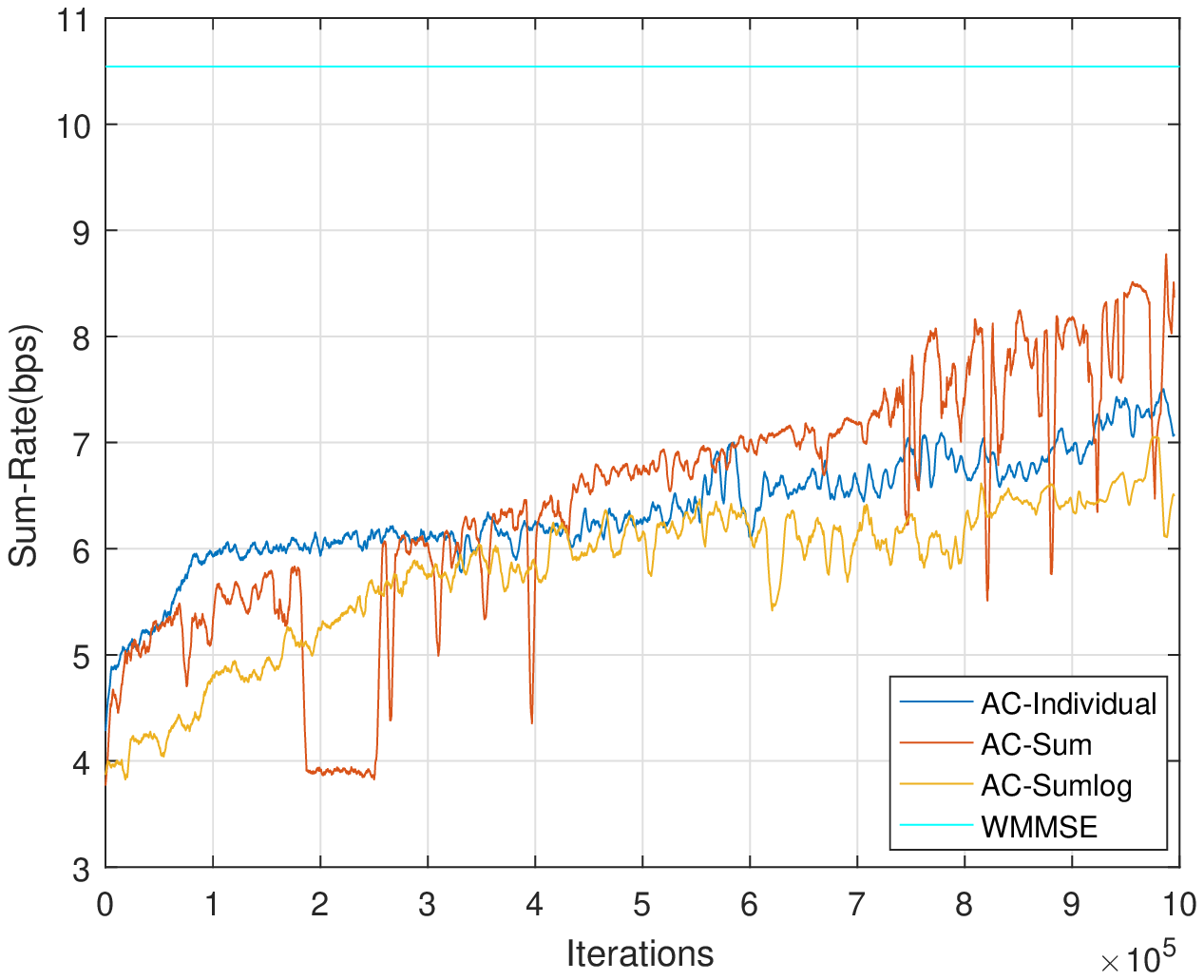}
	
	\caption{Moving average sum-rate achieved with three different reward mechanisms in the proposed actor-critic algorithm.}
	\label{ACsum}
\end{figure}

\begin{figure}[!t]
	\centering
	\includegraphics[width=3.5in]{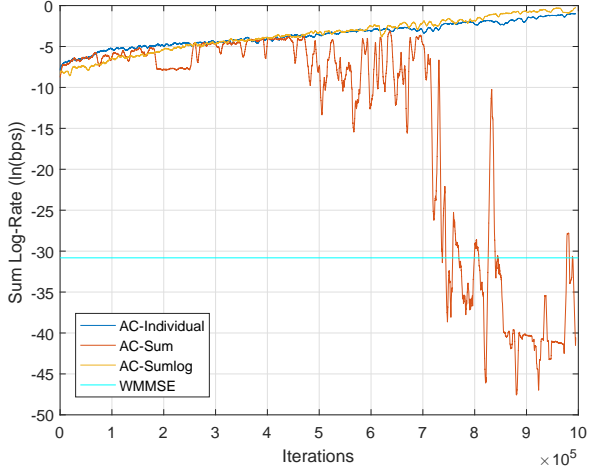}
	
	\caption{Moving average sum log-rate achieved with three different reward mechanisms in the proposed actor-critic algorithm.}
	\label{ACsumlog}
\end{figure}

Simulation results with the actor-critic DRL agent are presented in Figs. \ref{ACsum} and \ref{ACsumlog}. In this case, we observe similar performance trends as discussed above while noting that slighty lower performance levels are achieved compared with the DQN agent.

\subsubsection{Benefits of Power Allocation}
\begin{figure}[!t]
\centering
\includegraphics[width=3.5in]{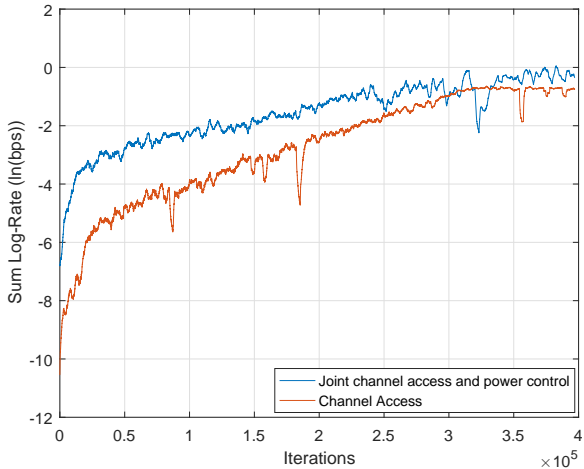}

\caption{Moving average sum log-rate with channel access only and joint channel access and power control. (DQN agent)}
\label{RvsBSumlog}
\end{figure}

\begin{figure}[!t]
\centering
\includegraphics[width=3.5in]{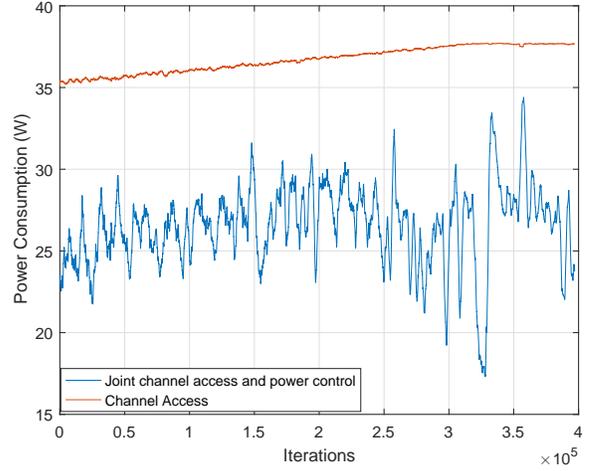}

\caption{Power consumption levels with channel access only and joint channel access and power control. (DQN agent)}
\label{Power}
\end{figure}

In this subsection, we demonstrate the benefit of introducing power allocation in addition to dynamic channel access. We note that power allocation plays an important role especially in dealing with fairness among the users. Most of the time, WMMSE algorithm results in a binary power allocation policy (0 or $P_{\max}$). On the other hand, when using individual SINR as the reward, all users may finally learn to transmit with full power because each user aims to maximize its own transmission rate ignoring the other users, hence in this case multi-level power allocation may not be needed. However, when fairness is considered among the users, some users with large direct gains tend to choose smaller transmit power to reduce the interference to the other users while maintaining a relatively high performance. With the DQN agent considered, the comparison of sum log-rate performance between joint channel access and power control and channel access only is shown in Figs. \ref{RvsBSumlog} and \ref{Power}. It is seen that when fairness is addressed, joint dynamic channel access and power allocation generally outperforms the channel access only, while using significantly less power.

\subsubsection{Testing in Dynamic Environment}
In this subsection, we conduct experiments again with both DQN and actor-critic agents, and analyze the performance in time-varying wireless channels. In this dynamic environment, the testing performance improves when more explorations are conducted during the training because the model will experience a larger number of different states. With this, we introduce the following modifications in the DRL agents.

\begin{itemize}
\item \emph{DQN}:
For DQN, we modify our $\epsilon$-greedy strategy during the training: $\epsilon$ decreases uniformly from 0.5 to 0.05 in the first half of the iterations and then is fixed to 0.05 for the rest of the iterations. We also deploy $\epsilon$-greedy policy with $\epsilon=0.1$ during the test. Such use of a small $\epsilon$ during the test is considered in order to enable limited explorations in the dynamic test environment with varying channel coefficients. With this, the algorithm is equipped with a capability to avoid getting stuck in a local optimum point.
\item \emph{Actor-Critic (AC)}:
 The best result in both training and testing achieved when $\epsilon$ decreases uniformly from 0.5 to 0.05 in the first 98\% of the iterations and then is fixed at 0.05 for the rest of the iterations.
\end{itemize}

\begin{figure}[!t]
\centering
\includegraphics[width=3.5in]{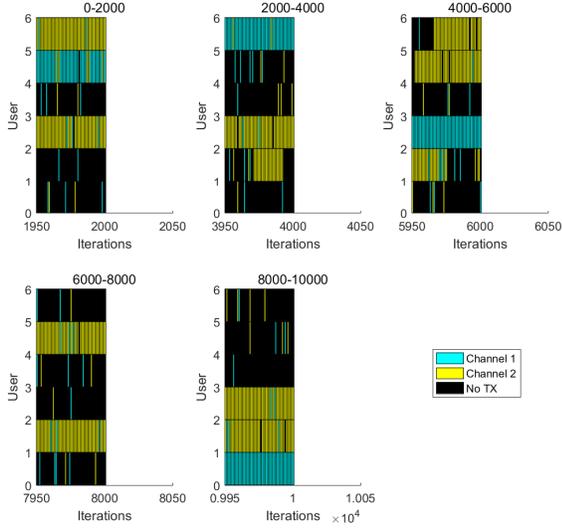}

\caption{Channel Selection Strategies of the DQN Agent in Testing.}
\label{DynamicCSCDRL}
\end{figure}

\begin{figure}[!t]
	\centering
	\includegraphics[width=3.5in]{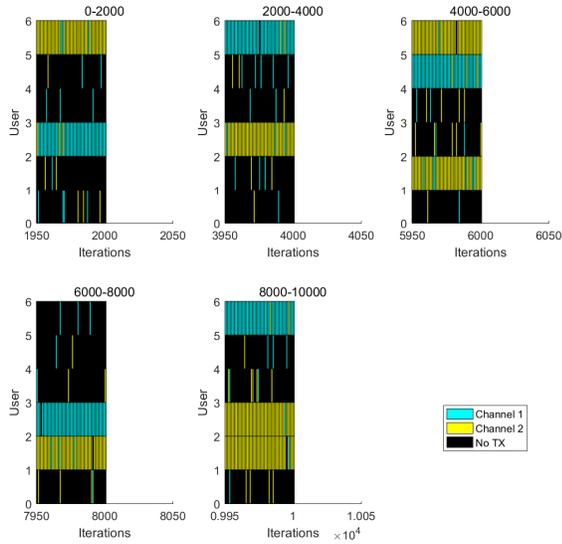}
	
	\caption{Channel Selection Strategies of Actor-critic Agent in Testing.}
	\label{DynamicCSCDRL_AC}
\end{figure}

\begin{figure}[!t]
\centering
\includegraphics[width=3.5in]{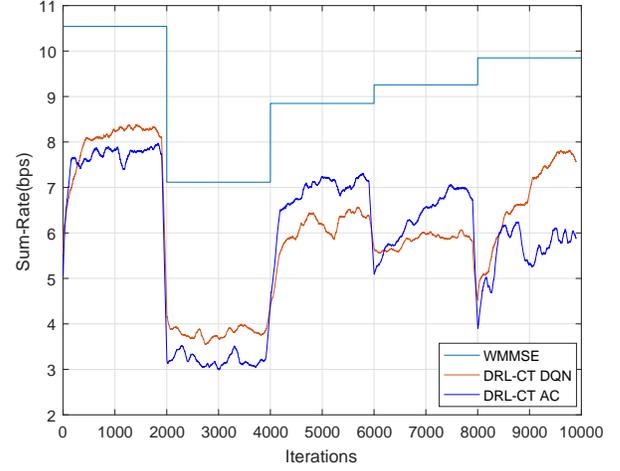}

\caption{Moving average sum rate achieved by CDRL duirng the testing.}
\label{DynamicSumCDRL}
\end{figure}

Once the DRL algorithms converge, they are tested in a dynamic channel described in Section \ref{subsec:dynamicchannel}. In these experiment, we set
%the time interval $T_{v}$ to 20ms,
the Doppler frequency $f_{d}$ to 15Hz.
It is further assumed that the channel is able to stay relatively stable in 2000 iterations before the next variation. For convenience, only the sum-rate maximization problem is discussed here. Figs. \ref{DynamicCSCDRL} and \ref{DynamicCSCDRL_AC} show the channel selection strategies of the users with DQN and actor-critic agents, respectively. The vertical axis indicates the 6 users and the horizontal axis indicates the iterations. We observe that the users are capable of learning transmission policies that are adaptive to dynamic channel conditions. Fig. \ref{DynamicSumCDRL} depicts the sum-rate performance achieved by both DQN and actor-critic agents along with the performance of the WMMSE algorithm. We see that as the channel varies, DQN and actor-critic agents, while exhibiting a certain performance loss with respect to the WMMSE approach, can adapt to the channel conditions  and learn relatively efficient transmission policies.

\section{Distributed and Federated Multi-Agent Deep Reinforcement Learning for Joint Channel Access and Power Control}
In this section, we address the case in which each user has its own DRL agent and even training is performed in a distributed fashion. We consider the same set of states, actions, and rewards as defined in Section IV. With these assumptions, each user builds its own DRL agent and is capable of learning locally, without uploading their transmission experiences during the training and downloading the global model afterwards.
%It is worth noting that in federated deep reinforcement learning (FDRL), each user learns based on an experience replay buffer that includes only its own experiences.
In the previous DRL framework in Section IV, training is centralized and is based on a buffer of experiences from all the users. In this section, we study distributed and also federated DRL (FDRL) training strategies. We focus on the sum-rate maximization problem, i.e. using the reward in (\ref{reward_sum}). First, we study distributed DRL (DDRL) and demonstrate that performance losses can be experienced in the test phase when operating in a dynamic channel environment. This will motivate us to employ federated learning policies to improve the performance. Hence, we subsequently focus on the proposed FDRL framework and examine its performance.

\subsection{Distributed Deep Reinforcement Learning (DDRL)}
%\textcolor{red}{Need decision on whether or not to add the actor-critic results.}\\
\subsubsection{Workflow}
In this case, we directly deploy the learning algorithm in Section IV to each user. The algorithm works as follows:
\begin{itemize}
\item \emph{Training}: Training is performed at the transmitters (user devices). Before training, each transmitter initializes its own DRL model. At the beginning of each time slot, each transmitter chooses a policy based on the $\epsilon$-greedy method and its local action policy. The following procedure is exactly the same as for the DRL with centralized training (DRL-CT) except that each user will store its own experience locally instead of uploading it to the central unit. And each user will update its own DRL model separately.
\item \emph{Usage}: When action policies of all the users converge, each user will make its decisions via its own DRL agent. The rest of the procedures is the same as the usage of DRL-CT.
\end{itemize}

\begin{figure}[!t]
\centering
\includegraphics[width=3.5in]{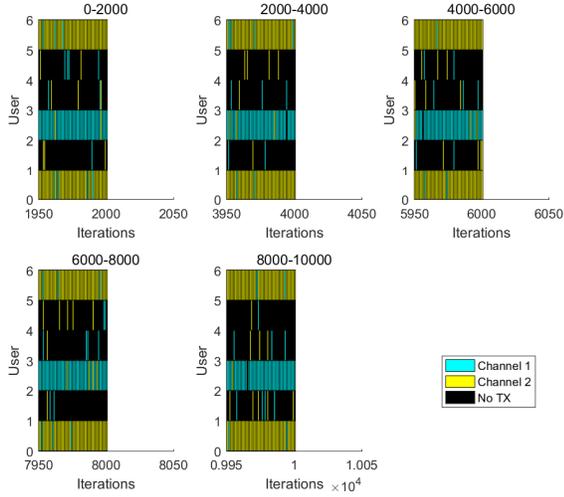}

\caption{Channel Selection Strategies of DDRL in testing. (DQN agent)}
\label{DynamicCSDDRL}
\end{figure}

\begin{figure}[!t]
\centering
\includegraphics[width=3.5in]{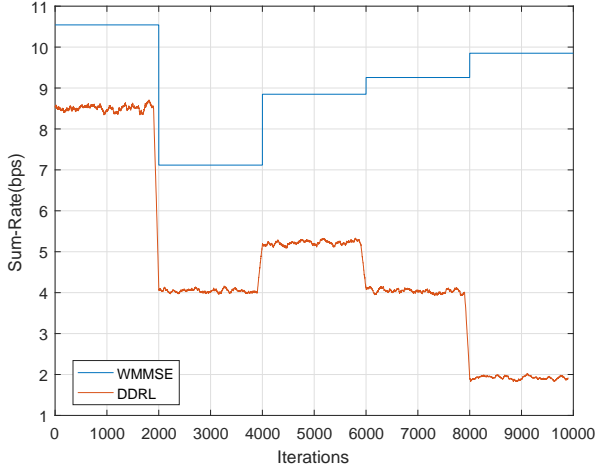}

\caption{Moving average sum rate achieved by DDRL during the testing. (DQN agent)}
\label{DynamicSumDDRL}
\end{figure}

We test the DDRL algorithm over dynamic channels with $f_{d}$ = 15Hz. Considering the DQN agent, users' channel selection strategies and sum-rate performances are depicted in Figs. \ref{DynamicCSDDRL} and \ref{DynamicSumDDRL}, respectively. Different from the case with DRL-CT, although users can still find a good policy for the original channel conditions, they experience difficulty in adapting their transmission strategies to the variations of the channel fading. Due to this, performance degradation is experienced as shown in Fig. \ref{DynamicSumDDRL}.

\subsection{Federated Deep Reinforcement Learning (FDRL)}
Potential degradation in the performance of DDRL in dynamic environments motivates us to apply federated averaging proposed in \cite{mcmahan2017communication} to deep reinforcement learning in a distributed fashion while imitating the process of centralized training. The algorithm of the proposed FDRL is provided in Algorithm \ref{Algorithm:FDRL} below.

\begin{algorithm}[!ht]
\caption{FDRL Algorithm}\label{Algorithm:FDRL}
\begin{spacing}{0.8}
\begin{algorithmic}[1]
\STATE{Each user calls the function \textbf{Initialization} of the corresponding DRL algorithm to initialize its own DRL model.}
\STATE{Randomly Select a group $\mathbb{G}$ consisting of $G$ users}
\FOR{time-slot $t = 1,...,T_{max}$}
\FOR{Each user $k \in \mathbb{G}$}
\STATE{All the users $\in \mathbb{G}$ make decisions simultaneously.}
\STATE{Each user $k$ will get access to its experience in the past $T$ time slots, i.e.
	$$s_k(t) = \{x_k(i)\}_{i=t-1-T}^{t-1}$$
where $x_{k}(i)$ denotes the observation of user $k$ in the time-slot $i$.}
\STATE{Each user calls the function \textbf{Action Selection} of their own DRL model to select an action $a$ and $\epsilon$-greedy method in the current state $s_{k}\left (t \right )$.}
\STATE{After the transmissions, each user will be given a feedback signal and a reward $r$ for action $a$ in state $s_{k}\left (t \right )$. }
\STATE{$s_{k}\left (t \right ) = s_{k}\left (t + 1 \right )$.}
\STATE{Each user records the $\{s_{k}\left (t \right ), a_{k}(t), r\left( t \right), s_{k}\left (t+1 \right ) \}$ tuple and calls the \textbf{Update} function of its own model to update the parameters.}
\ENDFOR
\FOR{every $T_{Fed}$ time slots}
\STATE{Each user $\in \mathbb{G}$ uploads the parameters of its DRL model to the center unit and a global action policy is obtained by averaging the parameters of the DRL models in group $\mathbb{G}$ at the center unit. }
\STATE{Randomly select another group $\mathbb{G{}'}$ of the users with replacement.}
\STATE{The global action policy is then sent back to the users in the new group $\mathbb{G{}'}$}
\STATE{$\mathbb{G}$ = $\mathbb{G{}'}$}
\ENDFOR
\ENDFOR
\end{algorithmic}
\end{spacing}
\end{algorithm}

We note that in Algorithm 5, the only difference from DDRL is the procedure of averaging the weights of all the DRL models in group $\mathbb{G}$ every $T_{Fed}$ time slots. Note that, when DQNs are used, all the parameters in the networks are uploaded and averaged. On the other hand, when actor-critic networks are considered, parameters of both networks will be uploaded and averaged in the central unit. In addition, since both the DQN and the actor neural network are action policy networks, it is assumed that they have the same structure and update policy when the federated learning process is enabled. Based on the discussion in \cite{mcmahan2017communication}, we can show that FDRL can ideally get the same performance as that of the DRL-CT by averaging the weights of the DRL models. Let us more specifically consider the case with DQN and assume that the users $\{1,2,\ldots, G\}$ with $G \le K$ are selected as group $\mathbb{G}$ in Algorithm 5. The DRL model parameters of user $k$ in time slot $t$ are denoted by $w_{t}^{k}$. If stochastic gradient descent (SGD) is used to update the weights $w_{t}^{k}$ of the action policy network, its gradient $g_{t}^{k}$ will depend only on the mini-batch $\mathbb{M}_{t}^{k}$ selected by user $k$ at time $t$, i.e. $g_{t}^{k} = \phi(\mathbb{M}_{t}^{k})$.

Without loss of generality, we can first assume $G=K$ and $T_{Fed} = 1$, which indicates that all $K$ users are selected in group $\mathbb{G}$ and their DRL models are averaged at every iteration after the local update. Then for user $k \left ( 1\leq k\leq K \right )$, we have
\begin{equation}
w_{t+1}^k = w_{t}^k - \alpha g_{t}^{k} = w_t - \alpha g_{t}^{k}.
\end{equation}
Assuming each $w_t^k$ is commonly initialized as $w_t$.

Parameters of the global action policy is then obtained by averaging the parameters of all $K$ deep neural networks, i.e.
\begin{equation}\label{FDRLupdate}
\begin{split}
w_{t+1} = \frac{1}{K}\sum_{k=1}^{K}w_{t+1}^{k} = \frac{1}{K}\sum_{k=1}^{K}(w_t-\alpha g_t^k)\\ = w_t - \frac{\alpha}{K}\sum_{k=1}^Kg_t^k
\end{split}
\end{equation}

Above, (\ref{FDRLupdate}) is equivalent to the update of parameters in DRL-CT as shown in (\ref{MiniBatch}), where $\mathbb{M} = \bigcup_{K=1}^{K}\mathbb{M}_k$.

Simulation results presented in the next section demonstrate that FDRL can successfully imitate the behavior of DRL-CT in both training and testing phases.
\subsection{Simulation Results}
\subsubsection{Simulation Setup}

In this section, SGD optimizer is used in the simulations instead of Adam optimizer, which is not applicable to FDRL. Additionally, several parameters are modified as listed in Table \ref{table2} below.
\begin{table}[!ht]
	\renewcommand{\arraystretch}{1.3}
	
	\caption{Experimental Parameters}
	\label{table2}
	\centering
	\small
	\begin{tabular}{|c||c|}
		\hline Exploration Time $\left(T_{e}\right)$ &500000\\
		
		\hline Learning rate $\left (\alpha \right )$&0.01\\
		
		\hline Discounted Factor $\left (\gamma \right )$ & $0.9$\\
		
		\hline Initial Exploration Probability $\left (\epsilon_{max} \right )$ &0.5\\
		
		\hline Final Exploration Probability $\left (\epsilon_{min} \right )$ &0.05\\
		\hline FDRL replay memory size & 100000\\
		\hline CDRL replay memory size & 600000\\
		\hline FDRL mini-batch size & 2\\
		\hline CDRL mini-batch size & 12\\
		\hline Optimizer &SGD\\
		\hline
	\end{tabular}
\end{table}

In the simulations, in addition to the FDRL with all $K = 6$ users in group $\mathbb{G}$,  we consider a more practical implementation of FDRL with $G=2$ and $T_{Fed}=100$, denoted as FDRL*. In FDRL*, $\epsilon$ of each user decreases asynchronously, which means that each user will start decreasing its $\epsilon$ only when it is selected in group $\mathbb{G}$ and starts training. Within the FDRL* algorithm, users take turns to contribute to the global model with their local data.

We also test the FDRL* algorithm with actor-critic agents. The actor and critic networks of each user are set to be the same as actor and critic networks of DRL with centralized training except the use of SGD optimizer and setting the learning rate of the actor network to be 0.001. During the training, each user decreases its $\epsilon$ from 0.2 to 0.05 with the exploration time $T_e$ as one-third of the total number of iterations in the training. As discussed before, each user trains and decreases $\epsilon$ when it is selected in group $\mathbb{G}$.

\begin{figure}[!t]
	\centering
	\includegraphics[width=3.5in]{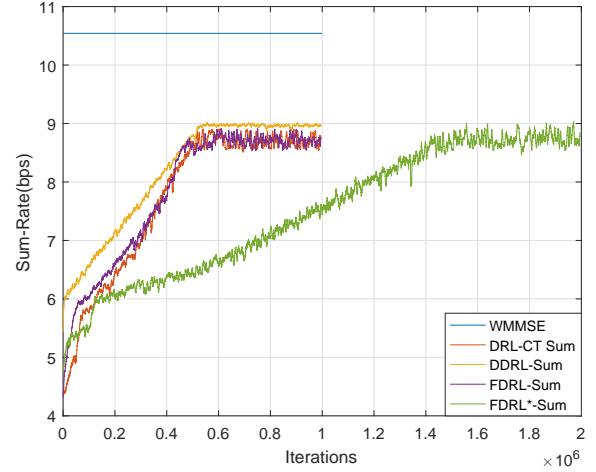}
	
	\caption{Moving average sum-rate achieved with different learning algorithms during the training. $K = 6$ user pairs and $N_c = 2$ channels.}
	\label{DynamicTraining}
\end{figure}

\begin{figure}[!t]
	\centering
	\includegraphics[width=3.5in]{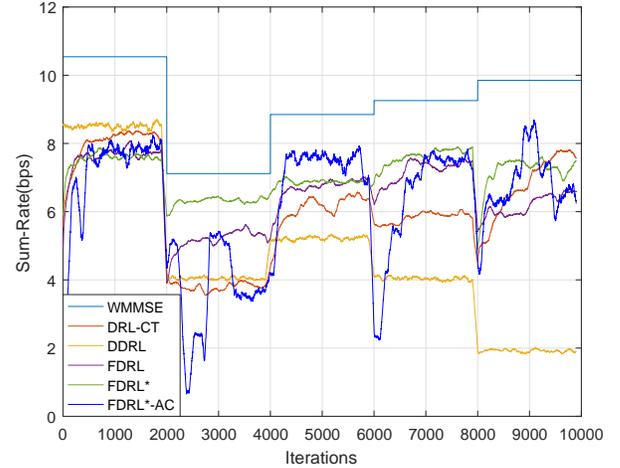}
	
	\caption{Moving average sum-rate achieved with different learning algorithms during the testing. $K = 6$ user pairs and $N_c = 2$ channels.}
	\label{DynamicTesting}
\end{figure}

\subsubsection{Sum-rate Performance}

Figs. \ref{DynamicTraining} and \ref{DynamicTesting} depict the sum-rate performance of the algorithms during training and testing, respectively. The overall training performance drops slightly compared to Fig. \ref{sum} due to the increase in the final value of $\epsilon$. It can be seen that although DDRL achieves the best performance during the training, it fails to adapt to the dynamic environment during the testing. DRL-CT, FDRL and FDRL* can achieve similar performance levels in training and all of them can learn an adaptive policy during the test. We take the average of 5 runs to obtain each curve in Fig. \ref{DynamicTesting} in order to alleviate the randomness during the test.

It is worth noting that both DRL-CT and FDRL take approximately $T_{e}$ iterations before convergence due to the selection of $\epsilon$. And FDRL* takes around three times the amount of time to converge. That is because only one third of the users are selected to train and decrease their $\epsilon$, and each user needs $T_{e}$ iterations to reduce its $\epsilon$ to $\epsilon_{min}$.

With the hyperparameters mentioned above, FDRL* with actor-critic agents can also learn an adaptive policy as shown in Fig. \ref{DynamicTesting}.
We note that using actor-critic agents has a non-negligible impact on the information exchange in the training phase. The amount of information exchange and computation cost in general increase due to the additional parameters introduced by the critic networks. However, as one advantage, there is no memory replay buffer required, which is preferable if the nodes have limited computational capacity and memory (e.g., in Internet of Things (IoT) networks with limited-capacity sensors and devices).

\begin{figure}[!h]
		\centering
		\includegraphics[width=3.5in]{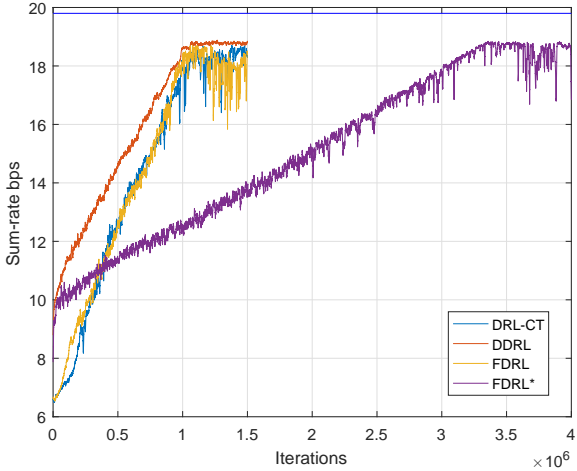}
		
		\caption{Moving average sum-rate achieved with different learning algorithms during the training. $K = 10$ user pairs and $N_c = 4$ channels.}
		\label{train}
	\end{figure}

\begin{figure}[!h]
		\centering
		\includegraphics[width=3.5in]{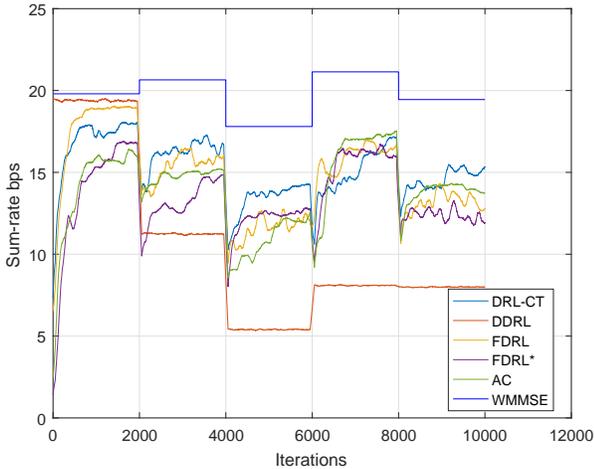}
		
		\caption{Moving average sum-rate achieved with different learning algorithms during the testing. $K = 10$ user pairs and $N_c = 4$ channels.}
		\label{test}
	\end{figure}

We have extended our simulation results to a larger network with $K = 10$ user pairs (i.e., 10 transmitters and 10 corresponding receivers) and $N_c = 4$ channels. Moving average sum rates in training and testing for different learning algorithms are plotted in Figs. \ref{train} and \ref{test}, respectively. In these figures, while larger sum rates are observed to be achieved, trends are similar to those in Figs. \ref{DynamicTraining} and \ref{DynamicTesting}.

\subsubsection{Analysis on Information Exchange}

We further compare the information exchange between DRL-CT and FDRL* to show that the federated averaging lower the communication burden.

\begin{figure}[!t]
	\centering
	\includegraphics[width=3.5in]{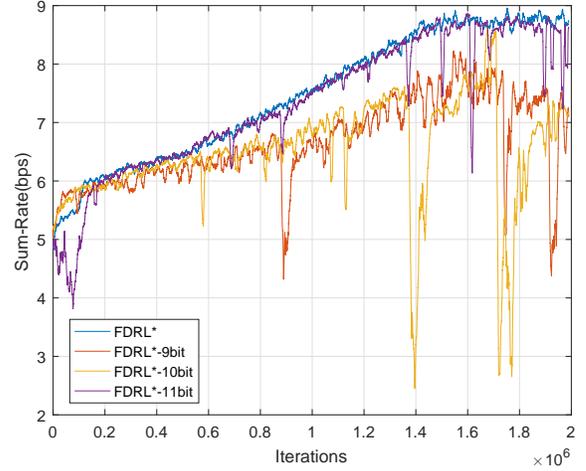}
	
	\caption{Moving average sum-rate achieved by FDRL* with different resolution of quantization}
	\label{Quantized}
\end{figure}

Sum-rate and individual SINR are the types of information exchanged in all the learning algorithms. The transition $\left ( s_{k}\left ( t \right ),a_k(t),r(t),s_{k}\left ( t + 1 \right ) \right )$ is uploaded to the central unit in DRL-CT at every iteration. On the other hand, parameters of the neural networks in group $\mathbb{G}$ are uploaded for averaging and downloaded afterwards at every $T_{Fed}$ iteration in FDRL*.
%It is further assumed that the downloading of the parameters in FDRL* from the central unit can be performed via broadcasting, and hence the downlink communication cost can be neglected.
We perform a low-precision quantization introduced in \cite{reisizadeh2020fedpaq} with different quantization levels and plot the corresponding sum-rate levels in Fig. \ref{Quantized}. We observe that the quantized version of FDRL* can get a comparable performance with the parameters quantized to 11-bit resolution or higher. We list the number of bits to be communicated in the training in Table \ref{table4} below.

\begin{table}[!ht]
	\renewcommand{\arraystretch}{1.3}
	
	\caption{Information Exchange (IE) during training}
	\label{table4}
	\centering
	\small
	\begin{tabular}{|c||c|}
		\hline Information & Number of bits \\
		\hline $s_k(t)$ & $ K \times \left ( 1+N_c+N_p+M \right )$ \\
		\hline a & 4 \\
		\hline $s_k(t+1)$ & $ K \times \left ( 1+N_c+N_p+M \right )$ \\
		\hline Parameters of each DQN & 3672 $\times$ 11 \\
		\hline
	\end{tabular}
\end{table}

It is worth noting that there is some analog information such as reward $r$ in the transitions of DRL-CT and a real-valued parameter extracted while performing the low-precision quantization for each user in FDRL*. In this section, since such analog information exchange is common to both cases, we only compare the digital parts of the information exchange during the training. For every 100 iterations, the communication rounds of DRL-CT and FDRL* is 100 and 1, respectively. Based on the above settings and assumptions, FDRL* requires only around 61$\%$ digital information exchange compared to that in centralized training for every 100 iterations.

\subsubsection{Training in Dynamic Environment}
Heretofore, we have analyzed the performance of the proposed DRL algorithms when they are trained in a static environment and are tested in a dynamic environment. The results so far show that, with the proposed framework, users are able to adapt their transmission strategies to the relatively slowly changing dynamic environments during the tests even if the training is conducted in a static environment. In this subsection, we further analyze how the proposed algorithms perform if the training also occurs in a dynamic environment. We first show that the Rayleigh fading with large but less-frequent variations  is equivalent to the case with slow but more-frequent variations. Subsequently, the experiments are conducted in different settings of the dynamic environment and the results are presented and analyzed.

Based on the description in Section \ref{subsec:dynamicchannel}, we can write $h_{ij}^{(n)(1)}$ and $h_{ij}^{(n)(2)}$ as
\begin{equation}
h_{ij}^{(n)(1)} = \rho h_{ij}^{(n)(0)} + e_{ij}^{(n)(0)} = \rho h_{ij}^{(n)(0)} + \sqrt{1-\rho^2}E_{ij}^{(n)(0)}
\end{equation}

\begin{equation}
\begin{split}
h_{ij}^{(n)(2)} = \rho h_{ij}^{(n)(1)} + e_{ij}^{(n)(1)} = \rho h_{ij}^{(n)(1)} + \sqrt{1-\rho^2}E_{ij}^{(n)(1)} \\= \rho (\rho h_{ij}^{(n)(0)} + \sqrt{1-\rho^2}E_{ij}^{(n)(0)}) + \sqrt{1-\rho^2}E_{ij}^{(n)(1)}\\ = \rho^2 h_{ij}^{(n)(0)} + \rho\sqrt{1-\rho^2}E_{ij}^{(n)(0)} + \sqrt{1-\rho^2}E_{ij}^{(n)(1)}\\ = \rho^2 h_{ij}^{(n)(0)} + \sqrt{1-\rho^4}E^*
\end{split}
\label{Rayleigh}
\end{equation}
where $E_{ij}^{(n)(t)}\sim \mathcal{CN}(0,1)$ and $E^* \sim \mathcal{CN}(0,1)$ are independent and identically distributed (i.i.d.) Gaussian random variables.

The formulation in (\ref{Rayleigh}) implies that the channel varying at every iteration with correlation $\rho$ is equivalent to a channel varying at every $T_v$ iterations with correlation $\rho^{T_v}$. It is assumed that the channel stays fixed in-between. We perform the experiments for different values of $T_v$, while maintaining the same intensity of channel variations for fair comparison.

\begin{figure}[!t]
	\centering
	\includegraphics[width=3.5in]{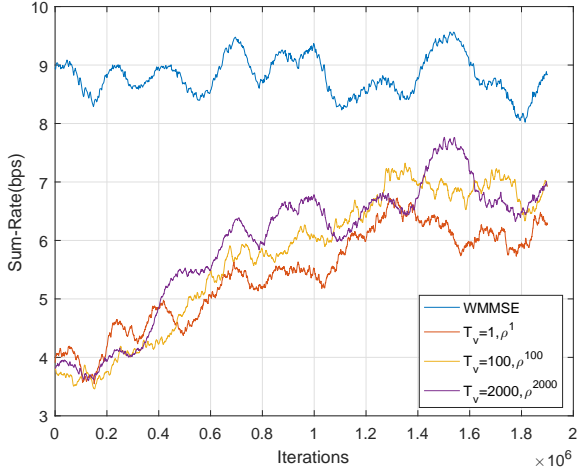}
	
	\caption{Moving average sum-rate achieved by DQN during the training under dynamic environment.}
	\label{SevereDynamicTesting}
\end{figure}

In Fig. \ref{SevereDynamicTesting}, we compare the training performance of DQN for different values of $T_v$ and their corresponding correlation $\rho^{T_v}$. Even if the variations are equivalent for different $T_v$, DQN achieves better performance as the value of $T_v$ increases.
We conclude that it is preferable to have longer duration of fixed channel conditions even though the variations are more significant when channel conditions eventually change.

\section{Conclusion}
In this paper, we have proposed a multi-agent DRL framework for performing the dynamic channel access and power control in wireless interference networks. We have considered both value-based DQN and policy gradient actor-critic DRL algorithms. We have first deployed this learning framework with centralized training and the simulation results have demonstrated that the users in the wireless interference network can efficiently allocate their transmission resources (e.g. power and spectrum) with the assistance of the proposed framework and can achieve around 90\% of the optimal performance achieved by the WMMSE algorithm for power control and exhaustive search for dynamic channel access. Subsequently, learning is conducted in a distributed fashion at the users, each of which is assumed to employ a DRL agent. We have addressed fully distributed training and learning, and also then employ federated DRL to imitate the process of the DRL with centralized training, while ensuring the privacy of the users and reducing the information exchange during training. Simulation results have indicated that DDRL can experience performance degradations, and user collaboration through federated averaging enables the users to learn policies that adapt to dynamically changing channel conditions and improve the performance.

\bibliographystyle{IEEEtran}
\bibliography{Ref}

\begin{IEEEbiography}[{\includegraphics[width=1in,height=1.25in,clip,keepaspectratio]{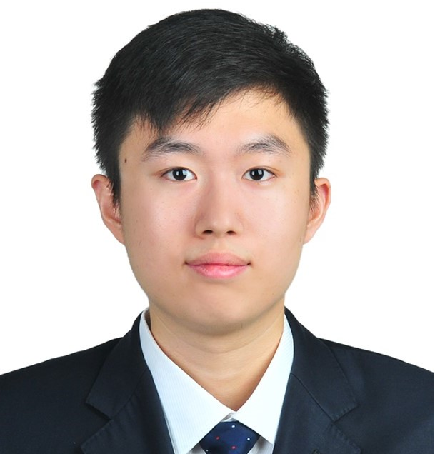}}]{Ziyang Lu}
	received the B.S. degree in electrical engineering from the University of Liverpool, U.K., in 2016, and the M.S. degree in electrical engineering from Syracuse University in 2018, where he is currently pursuing the Ph.D. degree with the Department of Electrical Engineering and Computer Science. His research interests are in the areas of wireless communication and machine learning.
\end{IEEEbiography}

\begin{IEEEbiography}[{\includegraphics[width=1in,height=1.25in,clip,keepaspectratio]{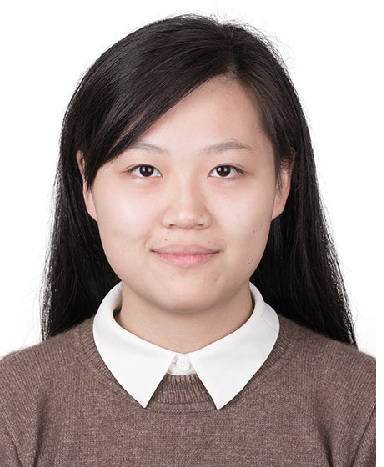}}]{Chen Zhong}
	 is currently a Ph.D. student in the Department of Electrical Engineering and Computer Science at Syracuse University. She received her B.S. degree in Information Engineering from Beijing Institute of Technology (China) in 2014 and her M.S. degree in Electrical Engineering from Stevens Institute of Technology in 2016. Her research interests are in the areas of wireless communication and networking. Currently, she is working on anomaly detection problems using tools from machine learning and optimization.
\end{IEEEbiography}

\begin{IEEEbiography}[{\includegraphics[width=1in,height=1.25in,clip,keepaspectratio]{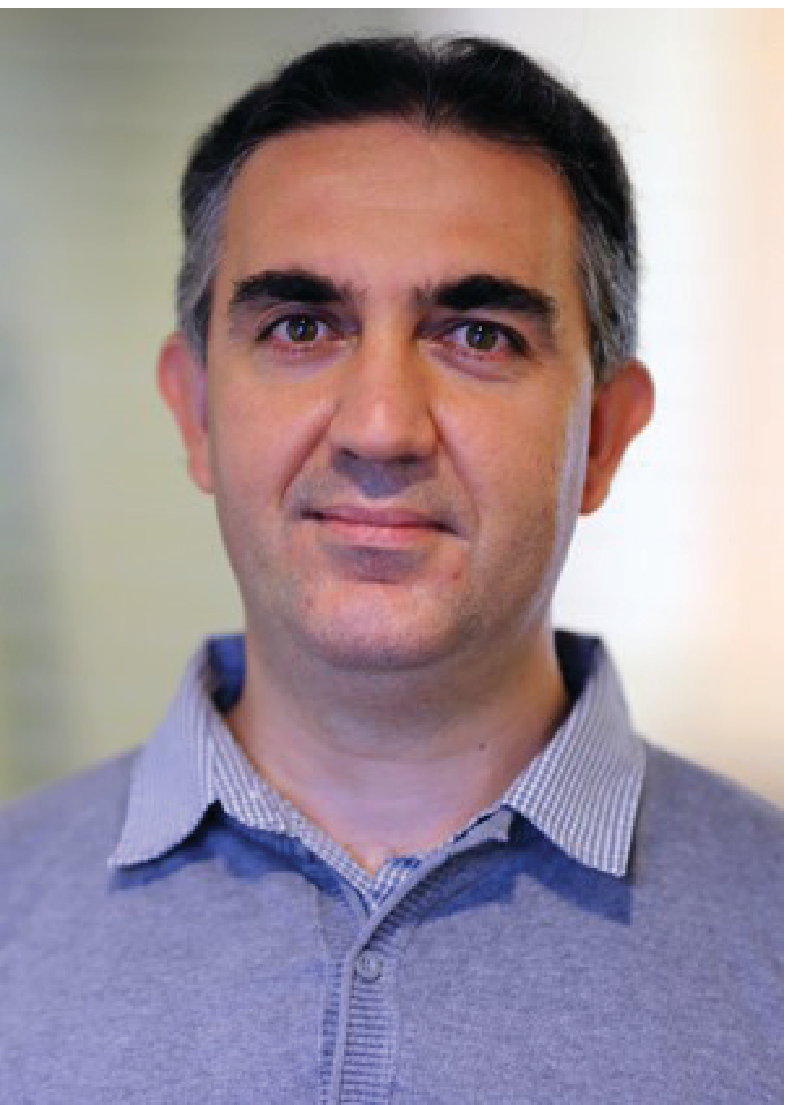}}]{M. Cenk Gursoy}
	received the B.S. degree with high distinction in electrical and electronics engineering from Bogazici University, Istanbul, Turkey, in 1999 and the Ph.D. degree in electrical engineering from Princeton University, Princeton, NJ, in 2004. He was a recipient of the Gordon Wu Graduate Fellowship from Princeton University between 1999 and 2003. He is currently a Professor in the Department of Electrical Engineering and Computer Science at Syracuse University. His research interests are in the general areas of wireless communications, information theory, communication networks, signal processing, and machine learning. He is a member of the editorial boards of IEEE Transactions on Wireless Communications, IEEE Transactions on Communications, and IEEE Transactions on Green Communications and Networking, and he is an Area Editor for IEEE Transactions on Vehicular Technology. He also served as an editor for IEEE Transactions on Wireless Communications between 2010 and 2015, IEEE Communications Letters between 2012 and 2014, IEEE Journal on Selected Areas in Communications - Series on Green Communications and Networking (JSAC-SGCN) between 2015 and 2016, Physical Communication (Elsevier) between 2010 and 2017, and IEEE Transactions on Communications between  2013 and 2018. He has been the co-chair of the 2017 International Conference on Computing, Networking and Communications (ICNC) - Communication QoS and System Modeling Symposium, the co-chair of 2019 IEEE Global Communications Conference (Globecom) - Wireless Communications Symposium, the co-chair of 2019 IEEE Vehicular Technology Conference Fall - Green Communications and Networks Track, and the co-chair of the 2021 IEEE Globecom - Signal Processing for Communications Symposium.  He received an NSF CAREER Award in 2006. More recently, he received the EURASIP Journal of Wireless Communications and Networking Best Paper Award, 2020 IEEE Region 1  Technological Innovation (Academic) Award, 2019 The 38th AIAA/IEEE Digital Avionics Systems Conference Best of Session (UTM-4) Award, 2017 IEEE PIMRC Best Paper Award, 2017 IEEE Green Communications \& Computing Technical Committee Best Journal Paper Award, UNL College Distinguished Teaching Award, and the Maude Hammond Fling Faculty Research Fellowship. He is a Senior Member of IEEE, and is the Aerospace/Communications/Signal Processing Chapter Co-Chair of IEEE Syracuse Section.
\end{IEEEbiography}

\end{document}